\documentclass[12pt]{article}
\usepackage{amssymb,float}
\begin{document}
\baselineskip 18pt
\title{Gravitational anomalies, gerbes, and hamiltonian quantization}
\author{{\bf C. Ekstrand, J. Mickelsson}\\Department of Theoretical Physics,
\\ Royal Institute of Technology,\\ SE-100 44 Stockholm, Sweden}
\date{}
\maketitle

\newcommand{\eq}{\begin{equation}}
\newcommand{\eqend}{\end{equation}}
\newcommand{\eqa}{\begin{eqnarray}}
\newcommand{\eqaend}{\end{eqnarray}}
\newcommand{\nonu}{\nonumber \\ \nopagebreak}
\newcommand{\Ref}[1]{(\ref{#1})}
\newcommand{\A}{{\cal A}}
\newcommand{\E}{{\cal E}}
\newcommand{\W}{{\cal W}}
\newcommand{\M}{{\cal M}}
\newcommand{\D}{{\cal D}}
\newcommand{\G}{{\cal G}}
\newtheorem{theorem}{Theorem}

\begin{abstract} In ref.\cite{CMM}, Schwinger terms in hamiltonian
quantization of chiral
fermions coupled to vector potentials were computed, using some ideas from
the theory of gerbes, with the help of
the family index theorem for a manifold with boundary. Here, we generalize
this method to include gravitational Schwinger terms.
\end{abstract}

\section{Introduction}
Chiral anomalies in quantum field theory appear in several different forms.
Historically they were first observed in perturbative calculations of certain
1-loop scattering amplitudes, as a breaking of the (classically valid)
chiral symmetry, \cite{ABJ}. Later nonperturbative methods were developed for
understanding the chiral symmetry breaking in the euclidean path integral
formalism, \cite{NRS},\cite{AW}. It was also understood that even the
symmetry under
coordinate transformations could be broken when quantizing massless fermions.

In the hamiltonian approach to chiral anomalies one considers the equal time
commutation relations for the infinitesimal generators of the classical
symmetry group. First one constructs the bundle of fermionic Fock spaces
parametrized by various external (classical) fields: gauge potentials,
metrics, scalar potentials etc. The quantization of the algebra of currents
in the Fock spaces requires some renormalization procedure. In $1+1$
space-time dimensions usually a normal ordering is sufficient but in
higher dimensions certain additional subtractions are needed.
Typically the renormalization modifies the algebra of classical symmetries
by the so-called Schwinger terms, \cite{S}.

In $1+1$ dimensions the Schwinger term is normally just a c-number, leading
to an affine Lie algebra (gauge transformations) or to the Virasoro
algebra (diffeomorphisms). In higher dimensions the algebra is more
complicated; instead of a central (c-number) extension the Schwinger
terms lead to an extension by an abelian ideal \cite{FM}.
Direct analytic computations
of the Schwinger terms in higher dimensions, although they can be carried
out \cite{M1} in the Yang-Mills case, are very complicated in the case
of an external gravitational field.
However, there are topological and geometrical
methods which give directly the structure of the quantized current algebra.
As in the case of euclidean path integral anomalies, a central ingredient in
this discussion is the families index theorem.

In a previous paper \cite{CMM} (see also \cite{CMM2} for a review) it was shown how
the Schwinger terms in the
gauge current algebra are related, via Atiyah-Patodi-Singer index theory,
to the structure of a system of local determinant line bundles
in odd dimensions. This system provides an example of a mathematical
structure called a bundle gerbe, \cite{Mu}.

In the present paper we want to extend the methods of \cite{CMM} for
constructing the Schwinger terms which arise in fermionic Fock space
quantization of the algebra of vector fields on an odd dimensional manifold.

In section 2 we set up the notation and recall some basic results in the
families index theory in case of compact manifolds with boundaries.
In section 3 we compute the curvature forms for a local system of complex
line bundles over a parameter space $B$ which consists of gauge potentials
and Riemannian metrics. In section 4 we explain how the Schwinger terms
in the Fock space quantization of vector fields (and infinitesimal
gauge transformations) are obtained from the local curvature formulas.
Finally, in section 5 we give some results of explicit computations of
the Schwinger terms.

\section{The family index theorem}

 Let $\tilde{\pi }:\tilde{M}\rightarrow B$ be a smooth fibre bundle
 with fibres diffeomorphic to a compact oriented spin manifold $M$ of even
 dimension $2n$. Assume that each fibre $M_z=\tilde{\pi }^{-1}(z)$,
 $z\in B,$ is equipped with a Riemannian metric.
 Assume further that $\tilde{M}$ is
 equipped with a connection. This means that at each point $x\in
 \tilde{M}$, the tangent space splits in a horizontal and a vertical
 part: $T_x\tilde{M}= H_x \oplus V_x,$ where $V_x$ consists of vectors
 tangential to the fibers.

 Let $\tilde{\E }$ be a vector bundle over $\tilde M$ which along each fiber
 $M_z$ is the tensor product ${\cal E}_z$ of the Dirac spinor bundle over $M_z$
 (with Clifford action of the vertical vectors $V_x$) and a finite dimensional
 vector bundle $W_z$ (with trivial Clifford multiplication).
It has a $\bf Z_2$ structure provided by the chirality operator
 $\Gamma$ according to: $c(\Gamma )=\pm 1$ on $\tilde{\E } ^{\pm }$,
 where $c$ denotes the Clifford action.

$\tilde{\E }$ is assumed to be
 equipped with a hermitean fiber metric. This naturally induces an
 $L^2$-metric in the space of sections  $\Gamma (M_z,{\cal E}_z)$.
Finally, let $D$ be an operator
 on $\Gamma (\tilde{M},\tilde{\cal E})$ which is fiberwise defined as a
 family of
 Dirac operators $D_z : \Gamma (M_z,{\cal E}_z)\rightarrow\Gamma
 (M_z,{\cal E}_z)$ with $z\in B.$
 With a Dirac operator we mean an operator
 that can be written as the sum of compositions of a covariant
 derivative and a Clifford multiplication.

 We would now like to apply the family index theorem to the case
described above. Before doing this, some additional assumptions is
needed. These are of different nature depending on if $M$ has a
non-empty boundary or not. When the boundary is empty, it will only be
assumed that the set $\{ D_z\} _{z\in B}$ should consist of
self-adjoint Dirac operators. The assumptions in the more difficult
case of a non-empty boundary $\partial M$ will now be described.

 We make the common simplifying assumption that for all $z\in B$, there
exists a collar neighbourhood of $\partial M_z$ such that all
structures in ${\cal E }_z$ are of 'product type\rq $ $ (see
ref. \cite{APS}). This implies that $D_z^{+}=D|_{\Gamma (M_z,{\cal E}
^+_z)}$ can be written as
$c_t(\frac{\partial}{\partial t}+D_z^{\partial })$ near the boundary,
where $c_t$
is the Clifford multiplication by an element corresponding to the
coordinate vector field $\frac{\partial}{\partial t}$ (which is along the
unit inward normal vector field at the boundary)
and $D_z^{\partial}$ is a self-adjoint Dirac operator on
${\tilde{\cal E } ^+}|_{\partial M_z}$. Our conventions are such that
$c_t$ is unitary.

 Let $\mbox{Ind}{D}_{\lambda }$, be the family
$\{\mbox{Ind}D_{z,\lambda }\} _{z\in U_{\lambda }}$, where
$\mbox{Ind}{D}=\mbox{ker} D^+\ominus \mbox{ker} D^-$ is the index
bundle in the sense of K-theory and $U_{\lambda }=\{ z\in B;\lambda
\notin \mbox{spec} \left( D_z^{\partial
}\right)\}$. The notation means that every operator $D_z^{+}$ will be
restricted to the domain $\{ \psi \in \Gamma ( M_z,{\cal E }
^+_z; P_{z,\lambda }\psi |_{\partial M_z}=0\}$, while $D_z^{-}$
will be restricted to $\{ \psi \in \Gamma ( M_z,{\cal E }
^-_z ; (1-P_{z,\lambda }) c_t \psi |_{\partial
M_z}=0\}$, where $ P_{z,\lambda }$ is the spectral projection of
$D_z^{\partial }$ corresponding to eigenvalues $\geq \lambda $. We
will assume that the Dirac operators $D_{z,\lambda }$ are
self-adjoint.

 The family index theorem has been proven in
\cite{AS2} when $\partial M=\emptyset$ and in \cite{BC}, based on
\cite{APS}, when
$\partial M\neq\emptyset$. It reads:
\eqa
\label{eq:FIT}
\mbox{ch}\left( \mbox{Ind}{D}\right) (z) & = & \int _{M_z}
\hat{A}(M_z)\mbox{ch}(W_z), \quad\partial M=\emptyset\nonu
\mbox{ch}\left( \mbox{Ind}D_{\lambda }\right) (z) & = & \int _{M_z}
\hat{A}(M_z)\mbox{ch}(W_z) - \frac{1}{2}\tilde{\eta }_{\lambda }(z),\,\,\, z\in U_{\lambda },\partial M\neq\emptyset, \nonumber
\eqaend
where
\eqa
\label{eq:AR}
\hat{A}(M_z) & = & \mbox{det}^{1/2}\left( \frac{iR_z/4\pi }
{\sinh (iR_z/4\pi)}\right) \nonu
\mbox{ch}(W_z) & = & \mbox{tr exp}\left( iF_z/2\pi\right) .
\eqaend
 We choose to not write down the definitions of the form $\tilde{\eta
}_{\lambda }$ or the curvature 2-forms $\tilde{R}$ and $\tilde{F}$ in
$\tilde{M}$, where $R_z=\tilde{R}|_{M_z} $ and
$F_z=\tilde{F}|_{M_z}$, since we will only need their explicit
expression in a simple special case. The only thing about the $\tilde\eta$ form
we need to know later is that it depends only on the boundary spectral data.
The zero degree part of $\tilde \eta$ is just the $\eta$-invariant of the
boundary Dirac operator.

 For $M=\emptyset$, the determinant line bundle DET is an object
closely related with the index bundle $\mbox{Ind}{D}$. It is a line
bundle over $B$, fibre-wise defined as $(\mbox{det ker} D_z^{+})^{\ast
}\otimes\mbox{det ker }D_z^{-}$. To define it globally over $B$ we
must also account for that the dimension of $\mbox{ker}D_z^{+}$ and
$\mbox{ker}D_z^{-}$ can jump as $z$ varies. For a detailed
construction, see \cite{BF}. For $M\neq\emptyset $ there exist a
similar construction of a bundle $\mbox{DET}_{\lambda }$ over
$U_{\lambda }$, closely related to $\mbox{Ind}D_{\lambda }$, see
ref. \cite{PZ}. In \cite{BF} and \cite{PZ} it has been shown that for
$\partial M=\emptyset$ and $\partial M\neq\emptyset$ there exists a
connection on DET and $\mbox{DET}_{\lambda }$, respectively, naturally
associated with the Quillen metric, \cite{Q1}, with curvature given by
\eqa
\label{eq:FD}
\frac{i}{2\pi} F^{\mbox{\footnotesize{DET}}}(z) & = & \left(\int _{M_z}
\hat{A}(M_z)\mbox{ch}(W_z) \right) _{[2]}, \quad  \partial M=\emptyset
\nonu \frac{i}{2\pi}F^{\mbox{\footnotesize{DET}}_{\lambda }}(z) & = &
\left(\int _{M_z} \hat{A}(M_z)\mbox{ch}(W_z) -
\frac{1}{2}\tilde{\eta }_{\lambda }(z)\right) _{[2]},\nonu
   && z\in U_{\lambda },\partial M\neq\emptyset ,
\eqaend
 where $\left[ 2\right]$ denotes the part that is a 2-form. Notice that the
2-form part of the right hand side of the family index theorem,
eq. \Ref{eq:FIT}, is equal to the right hand side of eq. \Ref{eq:FD}.

In the case of an odd dimensional manifold $N$  one can produce in a similar
way an element $\Omega\in H^3(B, \bf Z),$ by an integration over the fibers
$N_z,$
\eqa
\label{eq:FDD}
\Omega (z) & = & \left(\int _{N_z}
\hat{A}(N_z)\mbox{ch}(W_z) \right) _{[3]}, \quad  \partial N=\emptyset
\eqaend
where this time we pick up the component of the form which is of degree
$3$ in the tangential directions on the parameter space $B.$ This form
plays an important role in the hamiltonian quantization of external field
problems, \cite{CMM}. It is the Dixmier-Douady class of a gerbe. A
nonvanishing Dixmier-Douady class is an obstruction to quantizing chiral
fermions in a gauge invariant manner.

\section{Local line bundles over boundary \\ geometries}

Let $P$ be a fixed principal  $G$ bundle over $M$ with a projection $\pi:P
\to M$ and $FM$ the oriented
frame bundle of $M.$ The bundle $FM$ is a also a principal bundle, with
the structure group $GL_+(2n,\bf R),$ the group of real $2n\times 2n$
matrices with positive determinant.
Let $Q$ denote the product bundle $P\times FM.$
Let $B=\cal A \times \cal M$ where $\cal A$ is the affine space of
connections on $P$ and $\cal M$ is the space of Riemannian
metrics on $M.$ Locally, an element of $\cal A$ is written as a Lie$(G)$
valued 1-form on $M.$ We may view $Q$ as
a principal bundle $\tilde Q$ over $\tilde{M}=M\times B$ in a natural way,
as the pull-back under the projection $M\times B\to M.$

 With notations as in the previous section we define $M_z$ as the
manifold $M$ with metric given by $z\in B$.
Along the model fiber $M,$ let $\cal E$ be the tensor product of
the
Dirac spinor bundle and a vector bundle  $W$ over $M,$ the latter being  an
associated bundle to $P(M,G).$ We view
$\cal E$ in a natural way as a vector bundle $\tilde{\cal E}$ over $M\times B.$
Finally, we let $D_z:\Gamma (M_z,{\cal E
}_z)\rightarrow \Gamma (M_z,{\cal E }_z)$ be the Dirac operator
constructed from $z\in B$ in the usual way; in terms of a local coordinates
$A=A_{\mu} dx^{\mu}, \Gamma= \Gamma_{\mu} dx^{\mu},$ and with respect to  a
local orthonormal frame $\{e_{a}\}_{a =1}^{2n}$ of $TM_z$ we have
$$
D_{(A,\Gamma)}=\sum _{a,\mu =1}^{2n}\gamma ^{a}{e_a}^{\mu}
\left( \partial _{\mu }+ A_{\mu}+ \Gamma _{\mu }\right) ,
$$
where the ${e_a}^{\mu}$'s are the components of the basis vectors $e_a$ in
the coordinate frame and $\gamma^a$ is the Clifford multiplication by $e_a,$
with $\gamma^a\gamma^b +\gamma^b\gamma^a = 2\delta^{ab}.$

Let $\cal D$ be the group of orientation preserving diffeomorphims
of $M$ and $\cal G$ the group of gauge transformations in $P,$ i.e.,
the group of automorphims of $P(M,G)$ which projects to the identity
map on the base. The groups $\cal G$ and $\cal D$ act through pull-backs
on $\cal A$ and $\cal M.$ The group actions induce a fiber structure
in $B=\cal A \times \cal M$ but in order to obtain smooth moduli spaces
we restrict to the subgroups ${\cal G}_0\subset \cal G$ and ${\cal D}_0
\subset \cal D.$ The former is the group of based gauge transformations
$\phi$ leaving invariant some fixed base point $p_0\in P,$  $\phi(p_0) =p_0.$
The group ${\cal D}_0$ is defined as
$${\cal D}_0 =\{\phi\in{\cal D} | \phi(x_0) =x_0 \mbox{ and } T_{x_0}\phi=
\mbox{id} \}$$
for some fixed $x_0\in M.$ With these choices, we obtain the smooth
fiber bundles ${\cal M}\to {\cal M}/{\cal D}_0$ and ${\cal A} \to {\cal A}/
{\cal G}_0.$
This leads also to a fibering $B={\cal A}\times {\cal M} \to ({\cal A}\times
{\cal M})/({\cal D}_0\ltimes {\cal G}_0).$ Note that the group of symmetries
is the semidirect product of ${\cal D}_0$ and ${\cal G}_0$ since locally
an element of ${\cal G}_0$ is a $G$-valued function on $M$ and the
diffeomorphisms act on the argument of the function.

Following \cite{AS3}
we have a connection form $\omega$ on $\tilde Q=P\times FM\times B\to
M\times B$ which can be pushed forward to
a connection form on ${\tilde Q}/({\cal D}_0\ltimes {\cal G}_0).$ Along
$P\times FM$ the form $\omega$
is given by a connection $A\in \cal A$ and by the Levi-Civita
connection given by a metric $g\in \cal M.$ Restricted to the second
factor $B=\cal A\times \cal M$ the form $\omega$ is called the BRST ghost and
will be
denoted by $v.$ Since the total form $\omega$ should vanish along gauge
and diffeomorphism directions it follows that its value along these
directions in $B$ is uniquely  determined by the value of the
corresponding vector field on $P\times FM.$

In the case of gauge potentials, $B=\cal A,$ an infinitesimal
gauge transformation is given locally as a Lie$(G)$ valued function $Z$
on $M$ and then $v_{p,A}(Z_A)= Z(x)$ where $x=\pi(p)$ and $Z_A=\delta_Z A$ is
the vector field on $\cal A$ defined by the infinitesimal gauge transformation
$Z.$ In the case of
diffeomorphims, $Z$ is the ${\bf gl}(2n,\bf R)$ valued function defined
as the Jacobian (in local coordinates) of a vector field on $M.$ Again,
$v(Z_{\Gamma})$ is the 'tautological 1-form', $Z$ evaluated at a point $x\in M.$
The ghost $v$ in other directions is a nonlocal expression involving the
Green's function of a gauge covariant Laplacian, \cite{AS3, Do},
but we shall make
explicit use of $v$ only in the gauge and diffeomorphims directions.

Next let $N$ be an odd dimensional manifold without boundary and let
$M=[0,1] \times N,$ dim$M=2n.$ Given a principal bundle $P$ over $N$
we can extend it to a principal bundle over $M$ (to be denoted by the
same symbol $P$) by a pull-back defined by the projection $[0,1]\times N
\to N.$
We choose a fixed connection $A_0$ in the principal
bundle $P$ over $N$ and a fixed metric $g_0$ in $N.$
If $A$ is an arbitrary connection in $P$ we form a connection $A(t)$
(with $t\in [0,1]$) in the principal bundle $P$ over $M$ by
$$A(t) = (1-f(t)) A_0 + f(t) A,$$
where $f$ is a fixed smooth real valued function such that
$f(0)=0, f(1)=1,$
and all the derivatives of $f$ vanish at the end points $t=0,1.$
Similarly, any metric $g$ in $N$ defines a metric in $M$ such that along
$N$ directions it is given by
$$g_{ij}(t) = (1-f(t)) (g_0)_{ij}  + f(t) g_{ij}$$
and such that $\partial_t$ is a normalized vector field in $t$ direction,
orthogonal to the $N$ directions. However, for computations below it is
more convenient to use directly a homotopy connecting the the Levi-Civita
connection $\Gamma$ (constructed from the metric $g$) to the connection
$\Gamma_0$ (constructed from $g_0$). The formula for this homotopy is the
same as for the gauge potentials above.

We use the 'Russian formula', which is just an expression of the fact that
in a principal bundle with a connection the curvature form has \it
no components along fiber directions. \rm
The formula tells that when the total curvature
on $P\times FM\times{\cal A}\times {\cal M}$ is evaluated along vertical
directions in ${\cal A}\times {\cal M} \to ({\cal A}\times {\cal M})/
({\cal G}_0
\times {\cal D}_0)$ and along vector fields on $P\times FM$ the result is
$$ F^{\omega} = F^{A,\Gamma} = dA +\frac12 [A,A] + d\Gamma + \frac12 [\Gamma,
\Gamma],$$
where $\Gamma$ is the Levi-Civita connection.

Next we replace $\omega = A+\Gamma +v$ by the 'time' dependent connection
$$\omega(t) = (1-f(t)) (A_0 +\Gamma_0) +f(t) (A+\Gamma +v).$$
An evaluation of the curvature of a Dirac determinant bundle over $\cal A
\times \cal M$ involves an integration of a characteristic class $p_{n+1}
(F^{\omega(t)})$ over $M=[0,1]\times N$ and an evaluation of the $\tilde\eta$ form
on the boundary. Here $p_i$ denotes a generic homogeneous symmetric
invariant polynomial of degree
$i$ in the curvature.

Actually we have to restrict the construction of the determinant bundle
to subsets $U_{\lambda}$ in the parameter space $B$ on the boundary.
Here $U_{\lambda}$ is again the set of those points in $B$ such that the
associated
Dirac operator on $\partial M$ does not have the eigenvalue $\lambda.$
These sets
form an open cover of $B.$ In each of the sets $U_{\lambda}$ one can define
the $\tilde\eta$ form associated to Dirac operators $D_z -\lambda$ as a
continuous function of the parameters $z\in B.$

We shall restrict to the problem of determining the
curvature along gauge and diffeomorphism directions on the boundary. The $\tilde\eta$ form is
a spectral invariant and therefore the only term which contributes is the
appropriate characteristic class in the bulk $M.$

The integration of the index density in the bulk can be performed
in two steps. First one integrates over the time variable $t$ and then
the resulting expression is integrated over $N$ to produce a 2-form
on ${\cal A}(N) \times {\cal M}(N).$ All the computations involving the ghost
$v$ are restricted to vertical directions.

Restricting to the case of gauge potentials (calculations involving the
Levi-Civita connections are performed in the same way) we have
\eqa F_{ij}^{\omega(t)} &= &\partial_i A_j(t) -\partial_j A_i(t) +[A_i(t),
 A_j(t)]\nonu
F_{0i}^{\omega(t)} &= &f'(t) (A  -A_0)_i,\nonumber \eqaend

where we use the index $0$ for the $t$ component. For intermediate times
$0<t<1$ the curvature has components also to the vertical directions,
\eqa  (F^{\omega(t)})_{[0,2]}&= &\frac12 f(f-1)[v,v] \nonu
   (F^{\omega(t)})_{[1,1]}&=& v f'(t) dt + f(f -1) [A-A_0, v].\nonumber\eqaend

Here we have denoted by $(F)_{[i,j]}$ the component of a form $F$ which is
of degree $i$ in the tangential directions in $P$ and of degree $j$ in
the ghost $v.$

If $p_k$ is any homogeneous symmetric function of degree $k$ of the curvature
we set $p_k(A,F) = p_k(A,F,F,\dots,F)$ and then
\eqa \label{eq:POM} \int_M p_k(F^{\omega (t)})&=&
 k\int_N \int_{0}^{1} p_k(f'(t)(A+v-A_0),
F^{\omega(t)}) dt \nonu  &\equiv &\int_N \omega_{2k-1}(A+v,A_0).\eqaend

The form on the right, when expanded in powers of the ghost $v$, gives
forms of various degrees on the parameter space $B.$ We are interested in
the curvature form which is of degree 2 in $v.$ The degree zero part
gives just the Chern-Simons form $\omega_{2k-1}(A,A_0)$ and if $N$ were an
even dimensional manifold the degree 1 term would be the nonabelian
gauge anomaly. In low dimensions one gets familiar explicit formulas; as an
example, consider the case of a trivial bundle and $A_0 =0.$
When $n=1$ the relevant characteristic class is $p_2(F)=\frac{1}{2!}
(\frac{i}{2\pi})^2
\mbox{tr}\, F^2$ and the curvature $c_1$ along vector fields given by a pair $X,Y$
of infinitesimal gauge transformations on the one-dimensional manifold $N$
is
$$\frac{i}{2\pi}c_1(X,Y)=   \frac{1}{8\pi^2} \int_N \mbox{tr} \, A[X,Y]$$
and in the case $n=2,$ dim$ N=3,$ $p_3(F)=\frac{1}{3!}
(\frac{i}{2\pi})^3 \mbox{tr}\, F^3$, one gets
\eqa \frac{i}{2\pi} c_3(X,Y) & = & \frac{1}{48\pi^3} \int_N \mbox{tr} \,
\Big(  (AdA +dA \,A + A^3)[X,Y] \nonu
&&+XdA\, YA -YdA\,XA\Big) .\nonumber\eqaend
The case of Levi-Civita connection $\Gamma$ needs some extra remarks.
The reason is that we have actually two types of Chern-Simons forms (and
associated anomaly forms) depending whether we write the connection with
respect to a (local) orthonormal frame in the tangent bundle $TM$ or
with respect to the holonomic frame given by coordinate vector fields.

Formally, the two Chern-Simons forms (and associated polynomials in $v$)
look the same; they are given exactly by the same differential polynomials
in $\Gamma_{\mu\nu}^{\lambda}$ (coordinate basis) or in $\Gamma_{\mu a}^b$
(with respect to an anholonomic basis $e_a^{\mu}$). The difference is
(locally) an exterior derivative of a form (in $N$) of lower degree.
The difference $\Delta \omega$ of the Chern-Simons forms involves
the matrix function
$e_a^{\mu}(x)$ on $N.$ Since this function takes values in the group
$GL(2n-1,\bf R),$ which topologically is equivalent to $SO(2n-1),$
there might be a topological obstruction for writing $\Delta \omega$
globally as $d\theta$ for some form $\theta.$  The potential
obstruction is the winding number of the map $e: N\to GL(2n-1,\bf R),$
given by the (normalized) integral
$$w(e) = \frac{1}{(2n-1)!} \left(\frac{i}{2\pi}\right) ^{n+1}
\int_N \mbox{tr}\, (e^{-1} de)^{2n-1}.$$

The choice $\Gamma_{\mu b}^a$ in the anholonomic frame $e_a$ leads to
diffeomorphism invariant integral $\int_N \omega_{2n-1}(A,A_0)$
and there are no anomalies or 2-forms along Diff$(N)$ orbits in
$\cal M.$ On the other hand, there is a frame bundle anomaly related
to local frame rotations; this takes exactly the same form as the pure
gauge anomaly discussed above, \cite{BZ}.

The choice $\Gamma_{\mu\nu}^{\lambda}$ in the coordinate frame is
insensitive to the frame rotations $e_a \mapsto e'_a$ but it responds
to a local change of coordinates. Explicit formulas for the forms $c_{2n-1}$
along Diff$N$ orbits are given in the appendix.

\section{Schwinger terms from the local system of line bundles}

As we saw in the previous section, APS index theorem gives us a system
of local determinant bundles $\mbox{DET}_{\lambda}$ over certain open sets
$U_{\lambda} \subset B.$ The infinite-dimensional group ${\cal K} =
{\cal D}_0 \times {\cal G}_0$ acts in the parameter
space $B$ mapping each of the subsets $U_{\lambda}$ onto itself.
We denote $\bf k=$ Lie$(\cal K).$
In general, the determinant bundles are topologically nontrivial and
one cannot lift directly the action of $\cal K$ to the total space of
$\mbox{DET}_{\lambda}.$  Instead, there is a an extension $\hat {\cal K}$ which
acts in the determinant bundles.

The Lie algebra $\hat{\bf k}$ of $\hat{\cal K}$ is given in a standard way.
It consists of pairs $(X,\alpha),$ where $X\in {\bf k}$ and $\alpha$ is
a function on $B,$ with commutation relations
$$[(X,\alpha), (Y,\beta)] = ([X,Y], {\cal L}_X \beta -{\cal L}_Y\alpha
+ c(X,Y; \cdot) ),$$
where the Schwinger term $c$ is a purely imaginary function on
$B$ and
antisymmetric bilinear function on $\bf k.$ It is defined as the value
of the curvature of $\mbox{DET}_{\lambda}$ at the given point in $B$
along the
vector fields $X,Y$ on $B.$ Here ${\cal L}_X$ denotes the Lie
derivative on
$B$ along the vector field $X.$ The Jacobi indentity
in $\hat{\bf k}$ is
an immediate consequence of the fact that the curvature is a closed 2-form
on $U_{\lambda}\subset B.$

Let $U_{\lambda\lambda'}=U_{\lambda}\cap U_{\lambda'}.$ Over $U_{\lambda
\lambda'}$ there is a natural complex line bundle
$\mbox{DET}_{\lambda\lambda'}$
such that the fiber at a point $z$ is the top exterior power of the
finite-dimensional vector space spanned by all eigenvectors of the Dirac
operator $D_z$ on $N$  with eigenvalues in the range
$\lambda < \mu < \lambda'.$ If $\lambda' < \lambda,$ we set
$\mbox{DET}_{\lambda
\lambda'} = \mbox{DET}_{\lambda'\lambda}^*.$  By construction, we have a
natural isomorphism
$$\mbox{DET}_{\lambda\lambda'}\otimes \mbox{DET}_{\lambda'\lambda''}\simeq
\mbox{DET}_{\lambda \lambda''},$$
for all triples $\lambda,\lambda',\lambda''.$

\begin{theorem} (\cite{CMM})
For any pairs $\lambda,\lambda'$ of real
numbers one has
$$\mbox{\rm DET}_{\lambda\lambda'} \simeq \mbox{\rm DET}_{\lambda} \otimes \mbox{\rm DET}_{\lambda'}^*$$
over the set $U_{\lambda\lambda'}.$
\end{theorem}
Note that even though in \cite{CMM} the discussion was mainly around the
case of gauge potentials and gauge transformations, the proof of the
theorem was abstract and very general, not depending on the particular
type of parameter space for Dirac operators.

In the gerbe terminology the content of this theorem is that the gerbe
defined by the system of local line bundles $DET_{\lambda\lambda'}$ is trivial.
The line bundles can be pushed forward to give a family of local line bundles
on $B/\cal K$ since the spectral subspaces transform equivariantly under
gauge transformations and changes of coordinates. However, over $B/\cal K$
the gerbe is no more trivial, i.e., it cannot be given as tensor products
of local line bundles over the sets $pr(U_{\lambda}),$ where $pr: B \to
B/{\cal K}$ is the canonical projection. The obstruction to the trivialization
is an element of $H^3(B/{\cal K}, \bf Z),$ the Dixmier- Douady class of the
gerbe. In \cite{CMM} the DD class was computed from the index theory in the
case of Yang-Mills theory; the generalization to the case involving metrics
and diffeomorphism is straight-forward and the free part of the cohomology
class is given by the integral formula \Ref{eq:FDD}, with $B$ replaced by
$B/\cal K.$

The importance of the above theorem comes from the following simple observation.
Let $H_z=H_+(z,\lambda) \oplus H_-(z,\lambda)$ be the spectral decomposition
of the fermionic '1-particle' Hilbert space with respect to a spectral cut at $\lambda\in {\bf R}$, not in the spectrum of ${D_z}$. This determines a representation of the CAR algebra
in a Fock space ${\cal F}(z,\lambda),$ with a normalized vacuum vector
$|z,\lambda>.$ The defining property of this representation is that
$$a(u)|z,\lambda>=0= a^*(u')|z,\lambda>,$$
for $u\in H_+(z,\lambda)$
and $u'\in H_-(z,\lambda)$.
All creation operators $a^*(u)$ and annihilation operators $a(u)$
are anticommuting except
$$a^*(u) a(u') + a(u') a^*(u) = <u',u>,$$
where $<\cdot,\cdot>$ is the inner product in $H_z.$ If we change the vacuum
level from $\lambda$ to $\lambda'> \lambda,$ we have an isomorphism
${\cal F}(z,
\lambda) \to {\cal F}(z,\lambda')$ which is natural up to a multiplicative
phase. The phase is fixed by a choice of normalized eigenvectors
$u_1,u_2,\dots u_p$ in the energy range $\lambda < D_z < \lambda'$
and setting $|z,\lambda'> = a^*(u_1) \dots a^*(u_p)|z,\lambda'>.$
But this choice is exactly the same as choosing a (normalized) element in
$\mbox{DET}_{\lambda\lambda'}$ over the point $z\in B.$

Thus, setting ${\cal F}_z ={\cal F}(z,\lambda) \otimes
\mbox{DET}_{\lambda}(z)$ for
any $\lambda$ not in the spectrum of $D_z$ we obtain, according to
Theorem 1, a family of Fock spaces parametrized by points of $B$ but which
do not depend on the choice of $\lambda,$ \cite{M2}. This gives us a smooth
Fock bundle $\cal F$ over $B.$  The $\cal K$ action on the
 base  lifts  to
a $\hat{\cal K}$ action in $\cal F,$ the extension part in $\hat{\cal K}$
coming entirely from the action in the determinant bundles $\mbox{DET}_{\lambda}.$

The Schr\"odinger wave functions for quantized fermions in background fields
(parametrized by points of $B$) are sections of the Fock bundle.
It follows that the Schwinger terms for the infinitesimal generators
of $\hat{\cal K},$ acting on Schr\"odinger wave functions, are given by
the formula for $c$ which describes the curvature of the determinant bundle
in the $\cal K$ directions. In the case of $B=\cal A$ and
${\cal K}={\cal G},$
the elements in the Lie algebra are the Gauss law generators. This case
was discussed in detail in \cite{CMM}. More generally, we give explicit
formulas for the Schwinger terms in section \ref{sec:EC}.

\section{Explicit computations}
\label{sec:EC}
 The Schwinger term in $(2n-1)$-dimensional space will now be computed. This will be done by using notations for Yang-Mills, but it works for diffeomorphisms as well if a different symmetric invariant polynomial is used. Eq. \Ref{eq:POM} gives that
\eqa
&&\omega _{2n+1}(A+v,A_0)  =  (n+1)\int _0^1f^{\prime }p_{n+1}\Big( A+v-A_0 ,  \nonu && fdA+f^2A^2+(1-f)dA_0+ (1-f)^2A_0^2 -f(f-1)[A_0,A]\nonu &&
+f( f-1) [A-A_0,v]+\frac{1}{2}f(f-1)[v,v]\Big) dt,\nonumber
\eqaend
 The Schwinger term can be calculated from this expression. However, since we are only interested in the Schwinger term up to a coboundary, an alternative is to use the \lq triangle formula\rq $ $ as in \cite{MS}:
\[
\omega _{2n+1}(A+v,A_0)  \sim  \omega _{2n+1}(A_0+v,A_0)+\omega _{2n+1}(A+v,A_0+v),
\]

where \lq $\sim$\rq $ $ means equality up to a coboundary with respect to $d+\delta$, where $\delta$ is the BRST operator. This gives a simpler expression for the non-integrated Schwinger term and also for all other ghost degrees of $\omega _{2n+1}(A+v,A_0)$. Straight forward computations gives the result
\eqa
&& \omega _{2n+1}(A+v,A_0)_{(2)}  \sim  \frac{(n+1)n}{2}p_{n+1}\left( v, dv+[A_0,v],dA_0+A_0^2\right) \nonu
&& +\frac{(n+1)n(n-1)}{2}\int _0^1f^{\prime }(1-f)^2p_{n+1}\Big( A-A_0, dv+[A_0,v], \nonu && dv+[A_0,v] , fdA+f^2A^2+(1-f)dA_0\nonu && +(1-f)^2A_0^2- f(f-1)[A_0,A]\Big) dt\nonumber
\eqaend
where the index $(2)$ means the part of the form that is quadratic in the ghost. Inserting $n=1,2$ and $3$ gives:
\eqa
&& \omega _3(A+v,A_0)_{(2)}  \sim  p_2(v, dv+[A_0,v])\nonu
&& \omega _5(A+v,A_0)_{(2)}  \sim  3p_3\left( v, dv+[A_0,v],dA_0+A_0^2\right)
\nonu && +p_3(A-A_0, dv+[A_0,v], dv+[A_0,v])\nonu
&& \omega _7(A+v,A_0)_{(2)}  \sim  6p_4\left( v, dv+[A_0,v],dA_0+A_0^2,dA_0+A_0^2\right) \nonu
&& +p_4\big( A-A_0, dv+[A_0,v], dv+[A_0,v], \nonu
&& dA +\frac{2}{5}A^2+3dA_0+\frac{12}{5}A_0^2+\frac{3}{5}[A_0,A]\big) .\nonumber
\eqaend
This gives expressions for the non-integrated Schwinger term in a pure
Yang-Mills potential if $p_{n+1}$ is the symmetrized trace. The appropriate
polynomial to use for the Levi-Civita connection is
$p_{n+1}=\hat{A}(M)_{n+1}$,
according to eq. \Ref{eq:AR}. Using
\eqa
\hat{A}(M) & = & 1+\left(\frac{1}{4\pi }\right) ^2\frac{1}{12}\mbox{tr}\left( R^2\right) \nonu
&&+ \left(\frac{1}{4\pi }\right) ^4\left( \frac{1}{288}\left(
\mbox{tr}\left( R^2\right) \right) ^2 +
\frac{1}{360}\mbox{tr}\left( R^4\right)\right) + ...\nonumber
\eqaend
gives for $n=1$ and 2:
\eqa
\omega _3(\Gamma +v,\Gamma _0)_{(2)} & \sim & \left(\frac{1}{4\pi }\right) ^2\frac{1}{12}\left(vdv+2\Gamma _0v^2\right)\nonu
\omega _5(\Gamma +v,\Gamma _0)_{(2)} & \sim & 0.\nonumber
\eqaend

 Since the expression for $\omega _7$ is rather long we will omit to write it down. However, for the special case $\Gamma _0=0$ it becomes:
\eqa
&& \omega _7(\Gamma +v,0)_{(2)}  \sim  \nonu && \left(\frac{1}{4\pi }\right) ^4\left( \frac{1}{288}\cdot\frac{2}{3}\mbox{tr} \left( \Gamma dv\right)\mbox{tr}\left( dv\left( d\Gamma+\frac{2}{5}\Gamma ^2\right)\right) \right. \nonu &&+ \frac{1}{288}\cdot\frac{1}{3}\mbox{tr}\left( (dv)^2\right)
\mbox{tr}\left( \Gamma
\left( d\Gamma +\frac{2}{5}\Gamma ^2\right)\right) \nonu
&& \left. + \frac{1}{360}\cdot\frac{1}{3}\mbox{tr} \left(\left( R-\frac{3}{5}\Gamma ^2\right)
\left( (dv)^2\Gamma +(dv)\Gamma dv +\Gamma (dv)^2\right)\right)\right) .\nonumber
\eqaend
 This expression can be simplified if subtracting the coboundary
\eqa
&&\left(\frac{1}{4\pi }\right) ^4\frac{1}{288}\cdot\frac{2}{3}\left( \delta\left(\mbox{tr}\left( \Gamma dv\right)\mbox{tr}\left( \Gamma d\Gamma +\frac{4}{5}\Gamma ^3\right)\right)\right. \nonu && +\left. d\left(\mbox{tr}\left( v dv\right)\mbox{tr}\left( \Gamma d\Gamma +\frac{2}{3}\Gamma ^3\right)\right)\right) .\nonumber
\eqaend
 The result is
\eqa
&& \omega _7 (\Gamma +v,0)_{(2)}   \sim  \left(\frac{1}{4\pi }\right) ^4\left( \frac{1}{288}\mbox{tr} \left( v dv\right)\mbox{tr}R^2\right. \nonu
&& \left. + \frac{1}{360}\cdot\frac{1}{3}\mbox{tr} \left(\left( R-\frac{3}{5}\Gamma ^2\right)\left( (dv)^2\Gamma +(dv)\Gamma dv +\Gamma (dv)^2\right)\right)\right) .\nonumber
\eqaend

 The gravitational Schwinger terms are obtained by multiplying with the
 normalization factor $(i/2\pi )^{-1}$, inserting the integration over $N$ and
 evaluating on vector fields $X$ and $Y$ on $M$ generating diffeomorphisms.
 The Levi-Civita connection and curvature have components $(\Gamma )^{i^{\prime
 }}_{\,\, j^{\prime }}=\Gamma ^{i^{\prime }}_{i j^{\prime }}dx^i$ and $(R )^{
 i^{\prime }}_{\,\, j^{\prime }}=R ^{i^{\prime }}_{ij j^{\prime }}dx^i\wedge
 dx^j$. Recall that $(v(X))^{i^{\prime }}_{\,\, j^{\prime }}=\partial _{j^{
 \prime }}X^{i^{\prime }}$, see, for instance, \cite{BZ}.

 To illustrate how the Schwinger terms can be computed, we give the result for
 1 space dimension:
\[
-2\pi i\int _N\omega _3(\Gamma +v,\Gamma _0)_{(2)}(X,Y)  \sim -\frac{i}{48\pi }\int _N
\left(\partial _xX\right) \partial _x^2Ydx,
\]
where \lq $\sim$\rq $ $ now means equality up to a coboundary with respect to the BRST operator.

 When both a Yang-Mills field and gravity are present, the relevant polynomial is a sum of polynomial of type
\[
p_k\left( F^{\omega (t)}\right) \tilde{p}_l\left( R^{\omega (t)}\right) ,
\]
where the curvatures $F^{\omega (t)}$ and $R^{\omega (t)}$ are with respect to pure Yang-Mills and pure gravity, respective.
This gives 
\eqa
&& \int _Mp_k\left( F^{\omega (t)}\right) \tilde{p}_l\left( R^{\omega (t)}\right) \nonu
&=& \int _N\int _0^1\Big[ kp_k\left( f^{\prime }(t)(A+v_A-A_0), F^{\omega (f(t))}\right) \tilde{p}_l\left( R^{\omega (h(t))}\right)  \nonu
&& +lp_k\left( F^{\omega (f(t))}\right) \tilde{p}_l\left( h^{\prime }(t)(\Gamma +v_{\Gamma }-\Gamma _0),R^{\omega (h(t))}\right) \Big] dt\nonumber .
\eqaend
The expression is independent of $f$ and $h$ (see below). With a choice such that $f^{\prime }(t)=0$, $t\in [1/2,1]$ and  $h^{\prime }(t)=0$, $t\in [0,1/2]$, this implies that  
\eqa
\label{eq:REFD}
\int _Mp_k\left( F^{\omega (t)}\right) \tilde{p}_l\left( R^{\omega (t)}\right) & = &\int _N\left( \omega _{2k-1}(A+v_A,A_0)\tilde{p}_l(R_0)\right.\nonu &&\left. +p_k(F)\tilde{\omega }_{2l-1}(\Gamma +v_{\Gamma },\Gamma _0)\right) . 
\eqaend

 Thus, the Schwinger term in combined Yang-Mills and gravity is up to a coboundary equal to the part of the expansion of \Ref{eq:REFD} that is of second ghost degree. In particular, this implies that Schwinger terms which have one Yang-Mills ghost and on diffeomorphism ghost are in cohomology equal to the Schwinger term obtained from the form in \Ref{eq:REFD}. Thus, truly mixed Schwinger terms do not exist. Notice that if the background fields are vanishing then the Schwinger term is gravitational (although some parts of the form degrees are taken up by the Yang-Mills polynomial). This can give anomalies of Virasoro typ in higher dimensions. Observe that there is nothing special about gravity, a Yang-Mills Schwinger term is obtained by interchanging the role of $f$ and $h$. This does however not mean that the gravitational Schwinger term differ from the Yang-Mills Schwinger term by a coboundary. The terms with $k=0$ respective $l=0$ ruin this argument.

 It is easy to see that our method of computing the Schwinger term agrees with one of the most common approaches: The polynomial $p_k(F^n)-p_k(F_0^n)$ is written as $(d+\delta )$ on a form, the (non-integrated) Chern-Simons form. The Schwinger term is the given by the part of the Chern-Simons form that is quadratic in the ghost. For the case when both Yang-Mills and gravity are present, the relevant polynomial is a sum of polynomials 
\eq
\label{eq:POL}
p_k(F)\tilde{p}_l(R)-p_k(F_0)\tilde{p}_l(R_0).\nonumber
\eqend  

 There is an ambiguity in the definition of the Chern-Simons form; it is for instance possible to add forms of type $(d+\delta )\chi$ to it. However, an ambiguity of this type will only change the Schwinger term by a coboundary. It will now be shown that the ambiguity in the definition of the Chern-Simons form is only of this type. Thus, we must prove that closeness with respect to $(d+\delta )$ implies exactness. This can be done by introducing the degree 1 derivation $\triangle$ defined on the generators by: $\triangle (d+\delta )(A+v_A)=A+v_A, \triangle (d+\delta )(\Gamma +v_{\Gamma })=\Gamma +v_{\Gamma }, \triangle dA_0=A_0,\triangle d\Gamma _0=\Gamma _0$, and otherwise zero. Then $\triangle (d+\delta )+(d+\delta )\triangle $ is a degree 0 derivation which is equal to 1 on the generators. Therefore, if $\chi$ is closed with respect to $(d+\delta )$, then $\chi$ is proportional to $(d+\delta )\triangle \chi$. 

 An example of a (non-integrated) Chern-Simons form for the polynomial in \Ref{eq:POL} is 
\[
\omega _{2k-1}(A+v_A,A_0)\tilde{p}_l(R_0)+p_k(F)\tilde{\omega }_{2l-1}(\Gamma +v_{\Gamma },\Gamma _0).
\]
This is in complete agreement with \Ref{eq:REFD}.

\end{document}